# Development of a novel high-performance balanced homodyne detector

Hong Lin[+1], Mengmeng Liu[+1], Xiaomin Guo[*1,2], Yue Luo[1], Qiqi Wang[1], Zhijie Song[1], Yanqiang Guo[*1,2] and Liantuan Xiao[*1,2]

1 Key Laboratory of Advanced Transducers and Intelligent Control System, Ministry of Education, College of Physics, Taiyuan University of Technology, Taiyuan 030024, China

2 College of Physics, Taiyuan University of Technology, Taiyuan 030024, China

*guoxiaomin@tyut.edu.cn *guoyanqiang@tyut.edu.cn *xiaoliantuan@tyut.edu.cn

*Abstract*—True random numbers are extracted through measurements of vacuum fluctuations in quantum state components. We propose an improved scheme utilizing an optimization-based simulation methodology to enhance the temporal resolution of quantum state detection and processing efficiency of vacuum fluctuation signals in continuous-variable quantum random number generators (CV-QRNGs), while simultaneously maximizing the entropy content of quantum noise sources. This work presents the first application of optimization simulation methodology to balanced homodyne detector (BHD) circuit design, with particular emphasis on improving high-frequency transmission characteristics. The design framework prioritizes system stability and S-parameter sensitivity to optimize both circuit architecture and critical component parameters. The AC amplifier circuit was implemented through ADS high-frequency simulations using two ABA-52563 RF amplifiers in a cascaded configuration, with circuit modeling performed on Rogers 4350 substrate optimized for high-frequency applications. This approach enabled the development of a switched-configuration BHD featuring: 1) 1.9 GHz bandwidth, 2) 41.5 dB signal-to-noise ratio at 1.75 GHz, 3) 30 dB common-mode rejection ratio at 100 MHz, and 4) frequency response flatness within ±1.5 dB across 1.3-1.7 GHz. Additionally, the Husimi function is employed for entropy analysis to reconstruct vacuum state phase-space distributions, validating the detector's quantum measurement fidelity. The implemented system demonstrates a collective generation rate of 20.0504 Gbps across four parallel channels, with all output streams successfully passing NIST SP 800-22 statistical testing requirements.

*Index Terms*—Balanced homodyne detector，quantum random number generator，bandwidth，flatness，CMRR，QCNR

I. INTRODUCTION

The burgeoning research on Continuous Variable (CV) quantum states [1] has drawn significant attention in quantum information science. Information encoding and processing schemes utilizing Gaussian states (e.g., squeezed states, coherent states) through optical field quadrature measurements have demonstrated both theoretical promise and practical utility, with experimental validation reported in [2,3]. Squeezed states, serving as critical resources for quantum noise reduction, find prominent applications in precision measurement systems such as the Laser Interferometer Gravitational-Wave Observatory (LIGO) [4], where balanced homodyne detection enables precision measurement of Gaussian squeezed states to enhance detection sensitivity. Given the fragile nature of quantum correlations in squeezed states, detector implementations must precisely discriminate quantum noise from classical noise components (e.g., electronic or environmental noise) to preserve squeezed state characteristics. Consequently, BHD implementations for squeezed state characterization typically require a signal-to-noise ratio(SNR) exceeding 10 dB while operating effectively within kHz-MHz bandwidth regimes. Conversely, Gaussian-modulated coherent state applications (e.g., CV-QKD [5-10]) present distinct requirements. These implementations demand high-speed optical modulation/detection capabilities governed by sampling rates exceeding 100 MS/s and modulation bandwidths over 1 GHz, necessitating bandwidths spanning tens to hundreds of MHz. Unlike squeezed state detection, these systems prioritize simultaneous amplitude and phase quadrature detection with relaxed SNR requirements (typically 7-10 dB suffice). As a pivotal component in CV quantum systems, BHD performance metrics (SNR, bandwidth) require application-specific optimization. The divergent measurement requirements between squeezed state characterization and Gaussian-modulated coherent state detection underscore BHD's adaptability across the CV quantum technology spectrum.

Beyond conventional applications in squeezed state metrology and Gaussian-modulated coherent state systems, the evolving quantum information landscape has expanded BHD utilization to emerging quantum information tasks. Notable among these are quantum random number generation (QRNG) [11-13], particularly Continuous-Variable Quantum Random Number Generation (CV-QRNG) [14,15], which exhibits distinct advantages over discrete-variable (DV) approaches [16,17], including (i) Gbps-level generation rates, (ii) simplified hardware configurations with cost efficiency, (iii) enhanced environmental stability, (iv) improved quantum noise harvesting efficiency, and (v) inherent scalability. Quantum-derived randomness finds critical applications in modern communication systems [18], particularly for SNR characterization and bit error rate analysis [19], with additional implementations in electronic warfare [20] and radar countermeasure systems [21]. The theoretical upper bound for CV-QRNG generation rates is governed by $R_{max}=H_{min}\times f_s$, where $H_{min}$ represents quantum min-entropy and $f_s$ denotes the sampling rate. Existing enhancement strategies include: (i) local oscillator power optimization [22], (ii) inverse-squeezed component utilization [23], and (iii) adaptive sampling range control [24,25]. However, these approaches demonstrate limited entropy improvement. Our research group developed a novel parallel quantum frequency mode extraction scheme [26,27] that overcomes traditional serial architecture limitations through quantum/classical noise ratio optimization. Nevertheless, as parallel processing capabilities advance, the fundamental constraint reverts to entropy source bandwidth limitations. While vacuum fluctuations theoretically exhibit white noise characteristics across infinite bandwidth, practical implementations face Nyquist constraints. Sampling beyond twice the BHD bandwidth induces temporal correlations between samples, compromising randomness through autocorrelation effects. To maximize quantum entropy extraction efficiency, next-generation BHDs must satisfy three critical specifications: (i) >10 dB SNR for quantum noise dominance, (ii) GHz-level bandwidth for high-rate sampling, and (iii) <±1.5 dB flatness for uniform spectral response. These stringent requirements position BHD optimization as a pivotal research frontier in CV quantum technology development.

We present a comprehensive design methodology for high-performance balanced homodyne detection (BHD) systems optimized for continuous-variable quantum random number generation (CV-QRNG) applications. This work introduces, for the first time, an optimization-driven simulation framework for BHD circuit design, with a specific emphasis on enhancing high-frequency transmission characteristics. Our optimization process integrates system stability considerations with S-parameter sensitivity analysis to determine optimal circuit configurations and component parameters. Through Advanced Design System (ADS) simulations, we implemented a cascaded configuration using two ABA-52563 RF amplifiers, fabricated on Rogers 4350B substrate with an innovative switchable architecture for functional reconfiguration, yielding a high-performance BHD implementation through optimized cascade amplification. The implemented detector achieves: (i) 1.9 GHz bandwidth, (ii) 41.5 dB SNR at 1.75 GHz, (iii) 30 dB common-mode rejection ratio (CMRR) at 100 MHz, and (iv) frequency response flatness within ±1.5 dB across 1.3-1.7 GHz, satisfying the demanding specifications of contemporary CV quantum systems while enabling efficient entropy extraction. Our findings address fundamental BHD design challenges and establish a technological framework for developing secure, scalable, high-throughput QRNG systems.

II. TWO-STAGE CASCADE AMPLIFICATION BHD EQUIVALENT MODEL

Current BHD circuit architectures predominantly employ single-stage amplification configurations due to their inherent advantages in maintaining high gain and SNR through minimized component count and reduced electronic noise injection. The growing demand for BHDs in quantum random number generation systems has driven the need for enhanced bandwidth capabilities. Dual-stage amplification architectures offer a compelling alternative, enabling optimal gain-bandwidth distribution across stages and resulting in a superior overall gain-bandwidth product. However, the increased stage count introduces additional components, potentially degrading SNR through cumulative electronic noise. This architecture presents a fundamental trade-off between bandwidth enhancement and SNR preservation. To address this design challenge, we developed an equivalent circuit model for two-stage cascaded amplification, validated through comprehensive ADS simulations. Following rigorous performance evaluation and trade-off analysis, we implemented a dual-stage RF amplifier configuration as the core AC amplification module, achieving an optimal balance between bandwidth extension and noise performance.

Before component selection and optimization, we developed an equivalent circuit model (Fig. 1) for the BHD AC amplification stage. This modeling approach facilitates streamlined circuit analysis and design optimization, enabling comprehensive characterization of system behavior and performance metrics while improving design intuitiveness, accuracy, and implementation efficiency. This theoretical framework establishes the foundation for subsequent parametric optimization and design refinement.

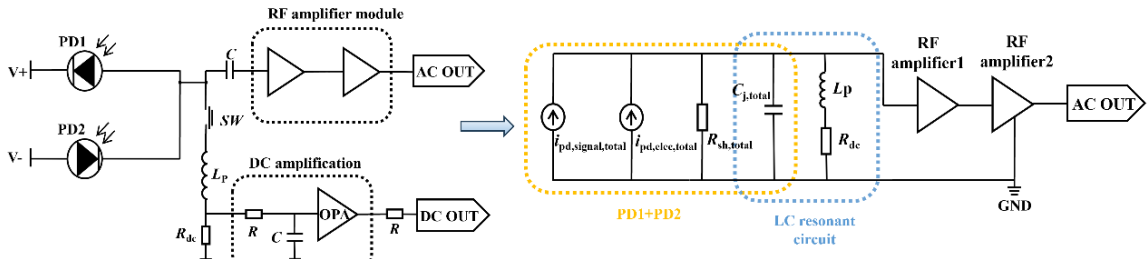

**Fig. 1.** Circuit Design Structure and AC Equivalent Model of BHD.

The photodiode pair (PD1 and PD2) in Fig. 1 operates in a series configuration, where the photocurrent is separated into DC and AC components via LC coupling. The AC component reflects quantum state quadrature fluctuations and inherent uncertainty. The equivalent circuit model (right side, Fig. 1) features a self-subtraction stage (yellow dashed box) representing the series photodiode pair, with $i_{PD,signal,total}$ denoting the combined RF noise, $i_{PD,elec,total}$ accounting for aggregate electronic noise, and $R_{sh,total}$ characterizing the shunt resistance. The LC resonant network (blue dashed box) integrates: (i) photodiode junction capacitance $C_j$, (ii) isolation inductor $L_p$, and (iii) sampling resistor $R_{dc}$, yielding the amplified AC output signal through dual-stage RF amplification.

Component selection was guided by comprehensive parameter analysis of photodiodes and RF amplifiers (Tables 1-2), ensuring optimal system performance.

TABLE 1

Parameters of different photodiodes

|  | LSIPD-A75 | LSIPD-A40 | LSIPD-LD50 | G8195 |
|---|---|---|---|---|
| Spectral range/nm | 800-1700 | 800-1700 | 800-1700 | 900-1700 |
| -3dB Bandwidth/GHz | 2.5 | 6.0 | 3.0 | 2 |
| Total capacitance/pF | 1.0 | 0.4 | 0.8 | 1 |
| Dark Current/pA | 18 | 20 | 5.0 | 20 |
| Photosensitivity/(A/w) | 0.90 | 0.85 | 0.90 | 0.95 |
| Shunt resistance/ GΩ | 50 | 30 | 100 | -- |
| brands | LIGHTSENSING | LIGHTSENSING | LIGHTSENSING | HAMAMATSU |

Photodiodes serve as the critical photoelectric conversion elements in balanced homodyne detection systems. Through comprehensive market analysis of commercially available photodiodes optimized for BHD applications, we identified four representative models (Table 2) for detailed performance evaluation. While enhanced bandwidth contributes to system performance, excessive bandwidth (e.g., 6 GHz) proves unnecessary. Instead, optimal performance requires careful consideration of: (i) dark current, (ii) junction capacitance, and (iii) shunt resistance, as these parameters fundamentally determine system-level performance. The LSIPD-LD50 demonstrates superior performance metrics including reduced dark current, minimized junction capacitance, and optimized shunt resistance, with optimized frequency response characteristics, where the -3 dB bandwidth selection aligns with theoretical SNR predictions [28].

$$R_{SN,PD_{OUT}} = 10 \lg \left( \frac{2e \cdot P \cdot S_\lambda}{\frac{4kT}{R_{sh}} + 2ei_{PD,dark}} \right), \tag{1}$$

Under 1.0 mW optical illumination, theoretical SNR values were calculated as 76.75 dB (LSIPD-A75), 75.93 dB (LSIPD-A40), and 82.13 dB (LSIPD-LD50), demonstrating LSIPD-LD50's superior SNR performance. Based on comprehensive performance evaluation, the LSIPD-LD50 pair was selected for differential photocurrent generation.

TABLE 2

Parameters of different RF amplifiers

|  | BGM1013 | BGA2817 | ABA-52563 |
|---|---|---|---|
| Bandwidth | 3GHz | 2.15GHz | 3.5GHz |
| Gain/dB | 35.5dB | 24.3dB | 21.5dB |
| Noise Figure/dB | 4.6dB | 3.9dB | 3.3dB |

The SNR in BHD is the ratio of quantum shot noise to electronics noise, which is calculated by the formula.

$$R_{SN} = 10 \lg \left( \frac{P_{signal}}{P_{elec}} \right), \tag{2}$$

Where $P_{elec}$ is the power of the BHD output electronic noise, $P_{signal}$ is the power of the BHD AC output signal. The SNR of the BHD developed in this paper is first predicted by the equivalent circuit model of the photodiode, and then the noise factor ($N_F$) of the cascaded RF amplifier, $N_F$, is utilized to calculate the SNR of the BHD, and the $N_F$ is calculated as

$$N_F(dB) = 10 \lg \left( \frac{R_{SN\_IN}}{R_{SN\_OUT}} \right) \approx 10 \lg \left( \frac{R_{SN,PD\_OUT}}{R_{SN\_OUT}} \right), \tag{3}$$

$$R_{SN\_OUT} = \frac{R_{SN,PD_{OUT}}}{10^{\frac{N_F}{10}}}, \tag{4}$$

Where $R_{SN\_IN}$ is the SNR at the input of the BHD and $R_{SN\_OUT}$ is the SNR at the output of the BHD. To facilitate the calculation, the SNR at the input of the BHD is regarded as the SNR at the output of the photodiode, i.e., $R_{SN,\ PDOUT} \approx R_{SN\_IN}$. In this paper the AC amplifier of the designed BHD is mostly cascaded with two RF amplifiers, then the total noise figure is:

$$N_F(\text{dB}) = N_{F1} + \frac{N_{F2} - 1}{G_1}, \tag{5}$$

where $N_{F1}$ and $N_{F2}$ represent the noise figures of the first and second amplification stages, with $G_1$ denoting the first-stage gain. Utilizing identical amplifier chips in both stages, we obtained total noise figures of 4.70 dB (BGM1013), 4.02 dB (BGA2817), and 3.41 dB (ABA-52563), as calculated from Table II parameters using Eq. (5). Applying Eq. (4), the theoretical output SNR values were determined as 27.83 dB (BGM1013), 32.55 dB (BGA2817), and 37.45 dB (ABA-52563), with corresponding performance characteristics illustrated in Fig. 2.

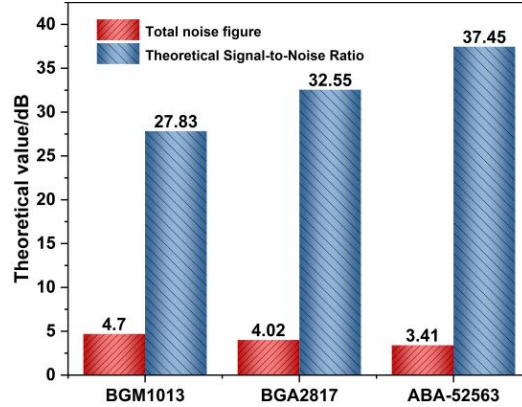

**Fig. 2.** Noise figure and theoretical output signal-to-noise ratio of each RF amplifier chip.

Considering critical performance metrics including 3 dB bandwidth, in-band gain flatness, and cost-effectiveness, we selected the ABA-52563 RF amplifier for its superior noise performance and optimized gain flatness characteristics, implementing a cascaded amplifier architecture to achieve high-gain AC signal amplification. Complementing this design, we implemented a DC amplification circuit using OP27 operational amplifiers in an isotropic proportional configuration.

The initial design phase established the fundamental BHD circuit architecture and corresponding equivalent circuit model, enabling preliminary theoretical analysis of noise characteristics and output SNR performance. However, theoretical modeling alone proves insufficient for optimal component selection. Therefore, we developed a comprehensive simulation framework for the AC amplification circuit, enabling optimized circuit design through advanced simulation analysis.

III. OPTIMIZED SIMULATION OF BHD SYSTEMS

Network modeling approaches characterize multi-port devices as black-box systems, enabling internal structure and performance analysis through frequency-dependent input-output characterization. This methodology facilitates network design and performance evaluation without requiring complete internal circuit knowledge. Furthermore, network simplification through input-output characteristic optimization reduces both active and passive component counts, thereby minimizing circuit complexity and implementation costs.

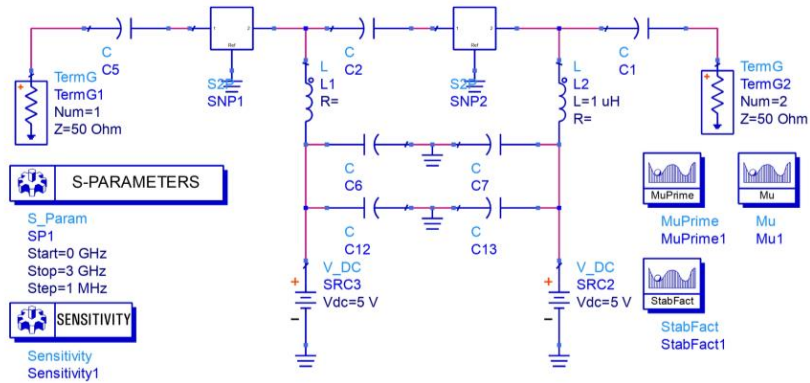

**Fig. 3** AC simulation link.

Our optimization methodology proceeds through the above stages: First, we model the AC amplifier circuit as a two-port network within the BHD system, establishing the simulation framework illustrated in Fig. 3. Component parameters are initialized based on RF amplifier datasheet specifications, followed by parameter optimization through iterative simulation analysis. Key components include: (i) AC coupling capacitors (C1-C5), (ii) bypass capacitors (C6-C7), (iii) power supply filtering capacitors (C12-C13), (iv) isolation inductors (L1-L2), (v) S-parameter models for ABA-52563 RF amplifiers (SNP1-SNP2), with 50 Ω terminations for both source (Term1) and load (Term2). Comprehensive analysis through ADS includes: (i) S-parameter characterization, (ii) stability evaluation, (iii) sensitivity analysis, (iv) parameter optimization, and (v) layout simulation. S-parameter analysis verifies signal transmission efficiency and impedance matching, stability evaluation confirms circuit operational stability, sensitivity analysis identifies critical components affecting performance metrics, parameter optimization achieves optimal design trade-offs, and layout simulation prevents performance degradation from substrate selection. This comprehensive simulation framework enables design optimization, and performance enhancement, and ensures robust, reliable BHD circuit implementation.

*3.1 Stability Simulation Analysis*

High-frequency circuit components inherently demonstrate parasitic capacitive and inductive effects that significantly influence circuit stability. When subjected to periodic voltage excitation, capacitors generate alternating currents, inducing electromotive forces in inductive elements, and creating feedback loops that potentially cause circuit oscillations. Consequently, comprehensive stability analysis is essential in high-frequency circuit design. Amplifier stability fundamentally determines the BHD's interference rejection and crosstalk suppression capabilities. We employ three stability metrics: (i) the stability factor K for overall amplifier stability assessment, (ii) the load stability factor Mu(R) for load impedance-specific stability evaluation, and (iii) the source stability factor μ(S)$\mu(S)$ for source impedance-dependent stability analysis. The circuit maintains stability when K>1, Mu(R)>1, and Mu(S)>1 [29]. Values below unity indicate compromised interference immunity, potentially causing signal oscillation and noise amplification.

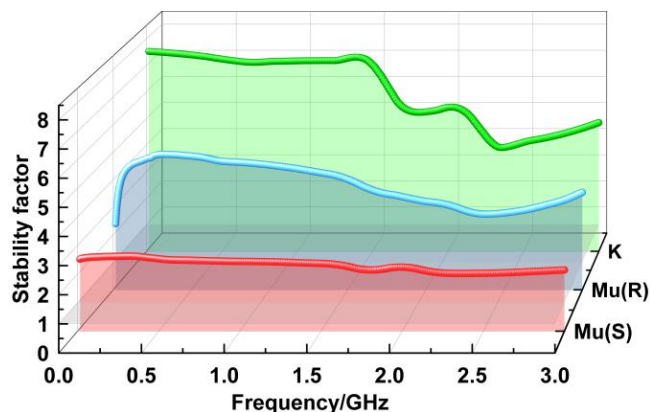

**Fig. 4.** Stability simulation results for AC link.

Figure 4 presents the simulated stability characteristics of the amplifier circuit, with the vertical axis representing stability factor magnitude across frequency. In RF circuit design, higher stability factor values correspond to enhanced circuit stability. The results demonstrate stable operation across the 0-3 GHz frequency range, with Mu(S), Mu(R), and K values are all greater than 1, confirming circuit stability within this bandwidth.

*3.2 Sensitivity Simulation Analysis*

Sensitivity analysis quantitatively evaluates circuit performance variation in response to parameter changes. This method identifies critical parameters influencing key performance metrics including gain, bandwidth, and efficiency, providing essential guidance for design optimization. Furthermore, it validates component compliance with design specifications.

Prior to optimization, we performed sensitivity analysis to identify critical components requiring optimization. Following S-parameter simulation requirements, we configured simulation targets using *GOAL* control, while analyzing coupling capacitors (C1, C2, C5) and isolation inductors (L1, L2) through Sensitivity control, with results presented in Figure 5.

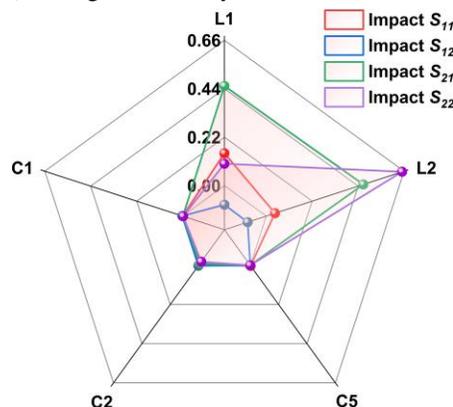

**Fig. 5.** Sensitivity analysis results for AC links.

The axes represent individual circuit components, with axis values quantifying parameter sensitivity effects on S-parameters. Higher absolute values indicate greater S-parameter sensitivity to component variations. The analysis reveals that isolation inductors L1 and L2 exhibit a dominant influence on S-parameters, with corresponding values quantifying their relative impact. Based on these findings, we will implement board-level experiments focusing on L1 and L2 optimization.

*3.3 Optimization Simulation*

Optimization simulation systematically adjusts circuit parameters to achieve predefined performance targets. We implemented parameter optimization using OPTIM control, establishing primary optimization targets to guide circuit optimization and performance enhancement. Initial simulations employed ideal models for capacitors and inductors, while actual components exhibit significant parasitic effects at high frequencies, potentially degrading circuit performance. To improve simulation accuracy, we implemented high-precision non-ideal models, specifically Murata component models, replacing ideal component representations.

Following S-parameter requirements, we configured optimization targets in *Goal* control, establishing: (i) $S_{11}$ and $S_{22}$ < -10 dB across 0-3 GHz, (ii) $S_{12}$ < -20 dB. Considering the ABA-52563's gain roll-off with frequency, we maintained $S_{21}$ > 40 dB at 3 GHz. We employed genetic algorithm optimization due to its superior convergence characteristics, computational efficiency, and parallel processing capabilities, enhancing simulation accuracy and reliability. Optimization results are presented in Figure 6.

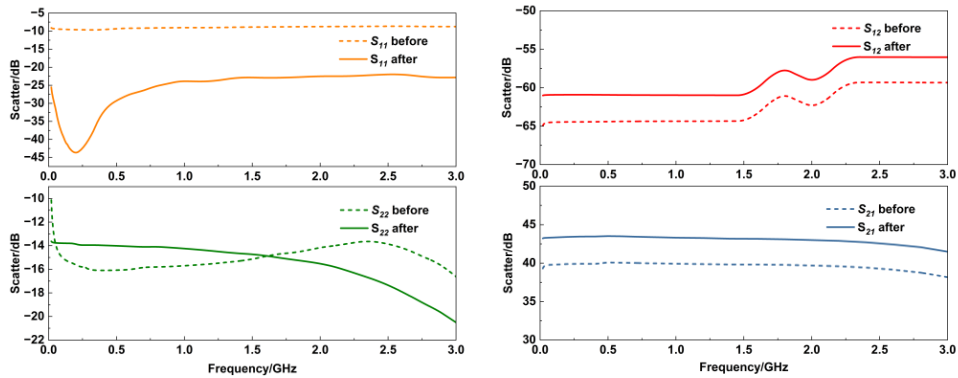

**Fig. 6.** Before and after optimization simulation results of AC link.

Pre-optimization results showed $S_{11}$ exceeding -10 dB across the frequency range and $S_{21}$ below 40 dB, failing to meet design specifications. Post-optimization achieved: (i) $S_{11}$ < -10 dB (0-3 GHz), (ii) $S_{22}$ < -10 dB, (iii) $S_{12}$ < -20 dB (0-3 GHz), and (iv) $S_{21}$ > 40 dB. All optimized parameters satisfied AC amplifier design requirements. These results validate the BHD amplifier circuit design methodology.

*3.4 Layout Simulation*

ADS Layout Simulation enables comprehensive circuit analysis through physical layout modeling within the ADS environment, facilitating layout verification and optimization using advanced ADS tools. This simulation methodology identifies potential issues arising from layout configurations, interconnects, and device packaging while supporting design optimization efforts. PCB substrate selection critically impacts circuit performance, with FR-4 and Rogers 4350 representing conventional and high-frequency substrate options. We conducted comparative layout simulations considering both substrate types. Simulation results are presented in Figure 7.

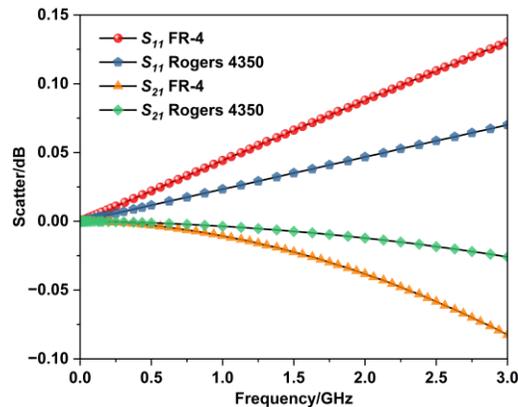

**Fig. 7.** EM simulation for FR-4 and Rogers 4350.

In RF/high-frequency design, minimizing the input reflection coefficient ($S_{11}$) is essential. $S_{11}$, expressed in dB, quantifies signal reflection magnitude. Higher $S_{11}$ values indicate increased signal reflection, demonstrating impedance mismatch and reduced transmission efficiency. Simulation results show FR-4 exhibiting higher $S_{11}$ values compared to Rogers 4350, confirming FR-4's inferior transmission characteristics.

$S_{21}$ characterizes signal transmission from Port 1 to Port 2, quantifying signal gain/attenuation through network components. Excessively negative $S_{21}$ values indicate substantial transmission losses, compromising practical system performance. FR-4 demonstrates inferior impedance matching compared to Rogers 4350, adversely impacting high-frequency signal transmission. Rogers 4350 offers superior dielectric properties, including reduced dielectric constant and loss tangent, minimizing high-frequency signal attenuation and distortion, thereby enhancing signal integrity and transmission efficiency. Furthermore, Rogers 4350 exhibits exceptional thermal stability and minimal thermal expansion, ensuring reliable high-frequency operation. Based on comprehensive theoretical analysis and simulation results, we selected Rogers 4350 as the optimal substrate for BHD implementation.

## IV. EXPERIMENTAL SETUP AND PERFORMANCE TESTING OF A BALANCED HOMODYNE DETECTION SYSTEM

We evaluated BHD performance through four key metrics including (i) detection bandwidth, (ii) SNR, (iii) frequency response flatness, and (iv) CMRR, using the experimental configuration illustrated in Figure 8. The device adopts a 1550 nm single-mode laser (ROI: LS-SM-1550) to output a continuous wave as the local oscillating light source of the balanced homodyne detection system, and adjusts the local oscillating optical power through a variable optical attenuator (VOA). In testing the CMRR, the supply voltage of the intensity modulator (IM) was set to half of its half-wave voltage to ensure that the modulation operated in its linear interval; at the same time, a sinusoidal signal was input to the intensity modulator via an arbitrary waveform generator (Hantek: HDG2102B). In the experiment, one input arm of the beam splitter (Beam splitter) was not fed with any signal and was in a vacuum state. After coupling by the 50:50 beam splitter, two optical signals with almost equal power were generated. According to Kirchhoff's current law, the photocurrent difference signals were formed by a series coupling of photodiodes. The BHD converted them into voltage signals, which were amplified and outputted to a spectrometer and an oscilloscope for further analysis in the frequency and time domains.

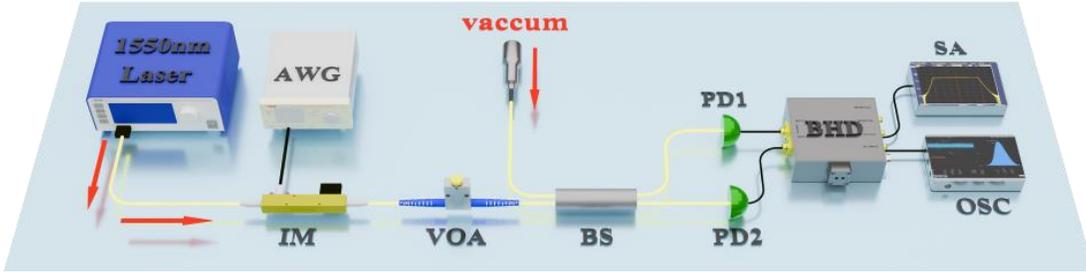

**Fig. 8.** Experimental configuration for BHD performance evaluation.

Ensuring electronic noise >10 dB below shot noise reference levels minimized measurement interference, maintaining SNR >10 dB. Our switch-enabled modular architecture separates DC and AC amplification paths, reducing DC circuit noise interference in AC measurements, and enhancing weak quantum signal detection capability. Switch closure enables DC amplification for incident light intensity monitoring. Following optimization simulations, we implemented 0.68 μH inductors (L1, L2) in the cascaded RF amplifier structure for experimental testing in a switch-disconnected configuration.

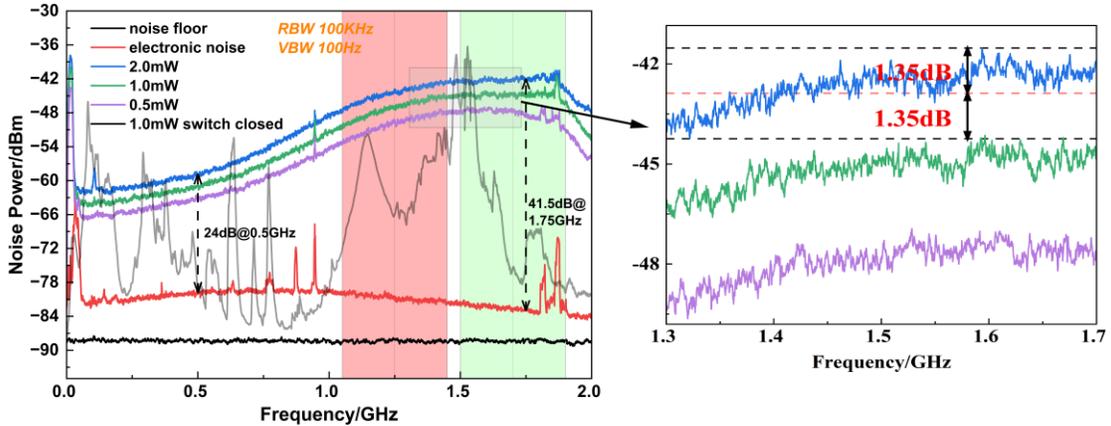

**Fig. 9.** BHD output spectra at various optical powers (switch-disconnected).

VOA-controlled optical power variation yielded the spectral characteristics shown in Figure 9. Experimental observations revealed saturation effects above 2.0 mW incident power, with diminishing spectral response indicating detector saturation. Systematic power variation from 0.5 mW to 2.0 mW demonstrated a 3 dB amplitude increase with optical power doubling. Spectrum analysis revealed 24 dB SNR at 500 MHz, attributed to 1/f and instrumental noise contributions, peaking at 41.5 dB

SNR (1.75 GHz), with ±1.35 dB flatness across 1.3-1.7 GHz, maintaining 38.5 dB SNR at 1.9 GHz (-3 dB bandwidth). Discrete frequency spikes in Figure 9 suggest external RF interference from environmental RF sources.

We further investigated SNR performance under different switch configurations, maintaining 1 mW incident power via VOA control, as illustrated in Figure 9. Switch-closed operation exhibited reduced SNR from DC circuit noise, demonstrating significant DC noise sensitivity. Switch-disconnected configuration eliminated DC branch interference, minimizing AC electronic noise, and yielding substantial SNR improvement. Switch-disconnected operation provided superior frequency flatness and SNR compared to closed-state operation.

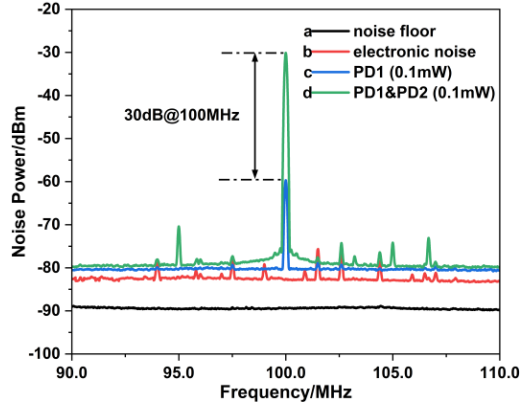

**Fig. 10.** Common mode rejection ratio results.

We quantified BHD performance through common-mode rejection ratio (CMRR) analysis, evaluating differential signal amplification and common-mode signal rejection capabilities, calculated from frequency-domain differential and common-mode signal measurements. For CMRR testing, we employed an arbitrary waveform generator (AWG) to produce a 100 MHz, 110 mV sinusoidal signal, driving an intensity modulator (IM) for optical intensity modulation. Experimental results are presented in Figure 10, with curve an indicating spectrometer baseline noise, curve b showing BHD electronic noise, curve c corresponding to single photodiode (PD1) illumination with PD2 blocked, and curve d representing simultaneous illumination of both photodiodes. This configuration achieved 30 dB CMRR at 100 MHz, satisfying design specifications.

We successfully implemented a BHD featuring: (i) 1.9 GHz bandwidth, (ii) 41.5 dB SNR at 1.75 GHz, (iii) 30 dB CMRR at 100 MHz, and (iv) <±1.5 dB flatness across 1.3-1.7 GHz.

V. HETERODYNE RECONSTRUCTION AND DETECTION FOR CV-QRNG

Husimi function reconstruction necessitates the simultaneous measurement of quantum state quadrature, achievable through heterodyne detection techniques, enabling Husimi function reconstruction from quadrature measurements. We implemented a 1550 nm local oscillator, with optical power and polarization controlled via VOA and PC. The optical signal passes through a 90° hybrid mixer, generating four phase-separated beams (0°, 90°, 180°, 270°), with 0°/180° and 90°/270° beams representing amplitude and phase components respectively. These beams are directed to two BHDs. A dual-channel signal generator provides 1.25 GHz and 1.7 GHz sinusoidal signals for mixing with BHD outputs. Low-pass filtering extracts 1.25 GHz and 1.7 GHz center frequencies, yielding 400 MHz bandwidth vacuum field sidebands at 1.25 GHz and 1.7 GHz as entropy sources. Path (a) routes signals to an oscilloscope for Husimi function reconstruction, while path (b) employs an 8-bit ADC sampling at 0.8 GS/s for random bit sequence generation, followed by FPGA-based post-processing through dual independent channels for final random number generation.

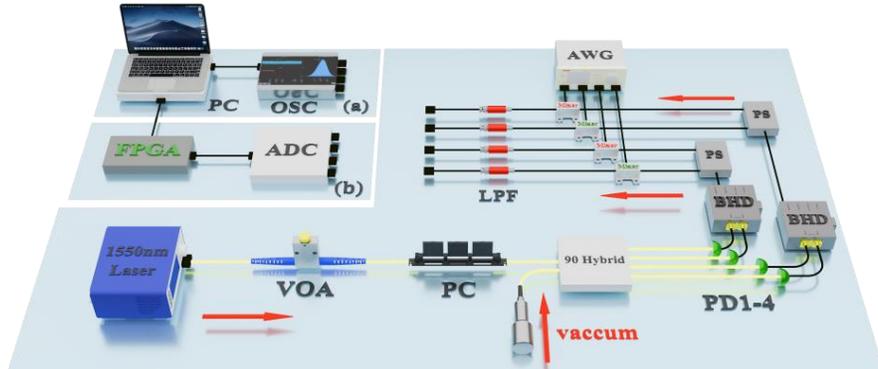

**Fig. 11.** Vacuum state Husimi quantum chromatography experimental setup.

*(a) Heterodyne reconfiguration*

System calibration before each experiment ensures measurement reliability. This procedure validates quantum measurement integrity while establishing phase space correspondence through detector output voltage calibration, enabling amplitude/phase variance calculation and phase space normalization. Using a VOA, we measured shot noise characteristics at optical powers from 0.2 to 1.8 mW in 0.2 mW increments, yielding P and Q quadrature variance measurements versus local oscillator power, as shown in Figure 12.

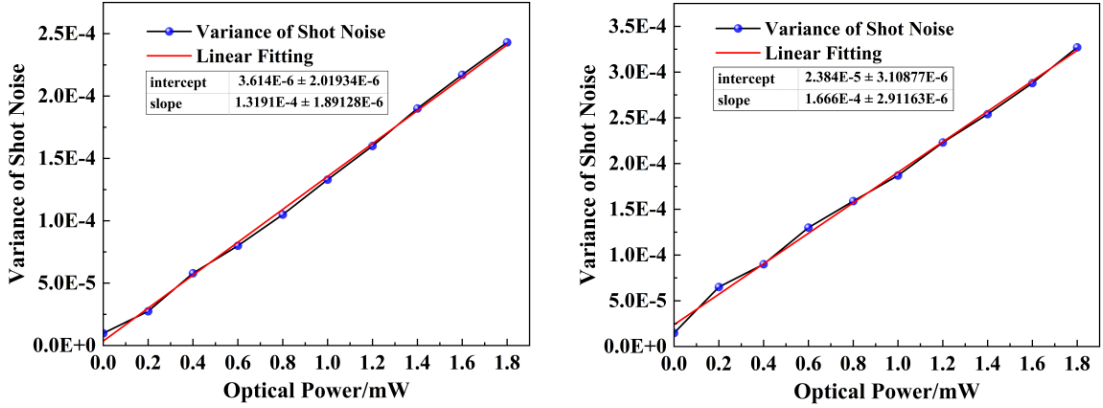

**Fig. 12.** Shot noise and linear fitting results for P(left panel) and Q (right panel) at different power.

Experimental reconstruction results are presented in Figure 13, showing excellent agreement between theoretical and experimental Husimi functions at 1.8 mW optical power, validating the performance of our custom BHD in the quantum random number generation system.

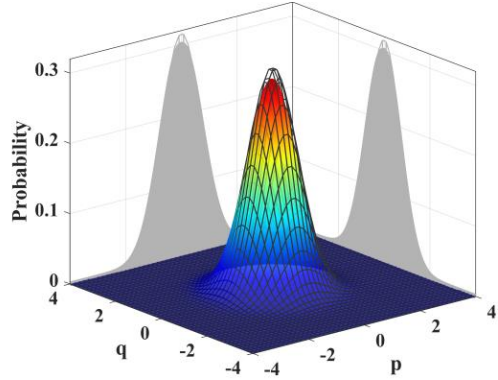

**Fig. 13.** Experimental vs. theoretical vacuum state Husimi function comparison.

*(b) Random number generation for differential beat detection and test results*

Our quantum random number generation scheme utilizes vacuum state fluctuations, where the random number generation rate depends on: (i) ADC sampling rate, (ii) ADC bit resolution, and (iii) minimum entropy. The sampling rate is determined by the BHD's effective bandwidth. Our custom BHD achieves 1.9 GHz effective bandwidth, yielding 400 MHz effective sampling bandwidth after mixing and low-pass filtering. We implemented 0.8 GS/s ADC sampling to prevent aliasing and signal distortion, with 8-bit resolution. Minimum entropy quantifies quantum state purity in raw random sequences, providing a conservative lower bound for entropy estimation. Raw data refers to the direct binary output from quantum measurement, containing both quantum noise and classical noise contributions. Post-processing extracts true quantum randomness from raw data. The minimum entropy per bit is defined as:

$$H_{min}(X) = -\log_2\left(max_{x\in\{0,1\}^n} P_X[X_i]\right), \qquad (6)$$

where n is the sampling precision of the ADC and $P_X$ is the probability distribution of the n-bit binary number X, i.e., the {0,1} distribution

The quantum conditional minimum entropy under classical noise E is:

$$H_{min}(M_{dis}|E) = -\log_2[\max(c_1, c_2)], \qquad (7)$$

where $c_1 = \frac{1}{2}\left[erf\left(\frac{e_{max}-R+\frac{3\delta}{2}}{\sqrt{2}\sigma_Q}\right) + 1\right]$, $c_2 = \left[erf(\frac{\delta}{2\sqrt{2}\sigma_Q})\right]$

R is half of the ADC input voltage range, $\delta = \frac{2R}{2^n}$, and the calculation is based on the irrelevance between quantum and classical noise. To simplify the analysis process and to ensure that the system operates under safe conditions with $c_1 \leq c_2$, finally, the quantum conditional minimum entropy can be simplified as

$$H_{min}(M_{dis}|E) = -\log_2\left[\text{erf}\left(\frac{\delta}{2\sqrt{2}\sigma_Q}\right)\right], \quad (8)$$

Utilizing the derived formula, we implement a random number extraction algorithm to obtain pure quantum random bits. The quantum-to-total entropy ratio directly influences the extractable quantum randomness and generation rate. Enhancing quantum noise entropy thus improves generation rates. System performance depends critically on the quantum-classical noise ratio (QCNR), with higher QCNR enabling greater true randomness extraction. Sufficient QCNR ensures random sequence quality and cryptographic security. We employ 1.25 GHz and 1.7 GHz center frequency sidebands (Figure 9) as entropy sources, achieving high QCNR for enhanced quantum random number generation.

True random number extraction rates are determined by Toeplitz matrix dimensions, governed by minimum entropy $H_{min}$ and security parameter $\varepsilon_{hash}$. FPGA implementation constraints hardware resource limitations, particularly logic elements. Following the Leftover Hash Lemma [30], post-processing extraction rates must satisfy specific constraints. Our implementation achieves 6.63 bits/8 bits (82.87%) minimum entropy, with FPGA resource optimization yielding $m=1729$ and $\varepsilon_{hash} = 10^{-50}$

$$m \leq n \times H_{min} - 2\log\frac{1}{\varepsilon_{hash}}, \quad (9)$$

The optimized Toeplitz matrix dimensions are 1729×2207, yielding 78.34% extraction efficiency, resulting in four-channel random number generation at 20.0504 Gbps.

We validated quantum random number performance through comprehensive testing of four-channel outputs against industry standards, including 15 NIST statistical tests using 375 MB test sequences at α=0.01 significance level. Combined four-channel outputs were similarly tested, yielding P-values >0.01 and pass rates within 0.9760-0.9946 confidence intervals (Figure 14).

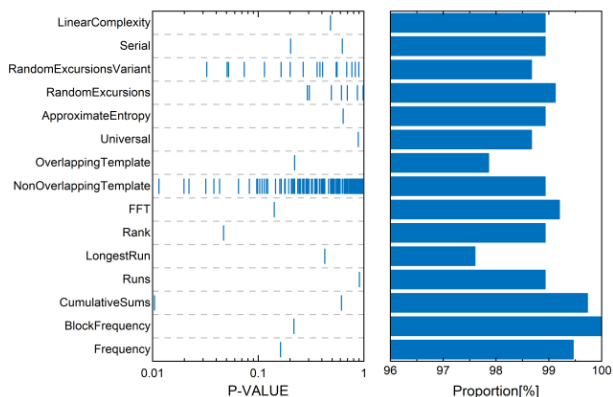

**Fig. 14.** Typical results of standard NIST Statistical Test Suite.

Additional validation employed TestU01 and Diehard test suites, with results presented in Figure 15. Our implementation successfully passed all statistical tests, validating the reliability and randomness of our parallel QRNG implementation.

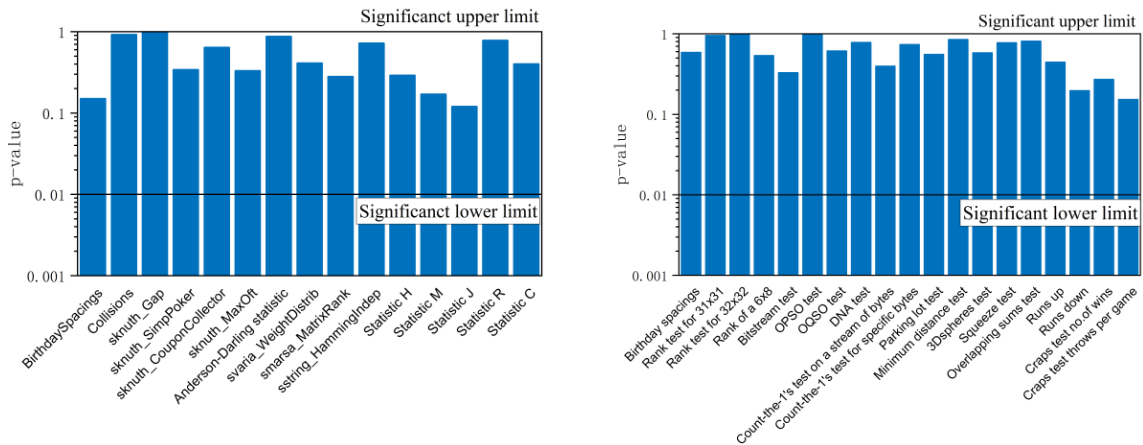

**Fig. 15.** (left) TestU01-SmallCrush results; (right) Diehard test results.

## VI. Summary

We present a high-performance BHD featuring wide bandwidth, high SNR, and stable in-band flatness, optimized for continuous-variable quantum key distribution (CV-QKD) and quantum random number generation (QRNG) applications. Initial theoretical analysis established an equivalent circuit model for noise characterization and SNR optimization. We developed a comprehensive simulation framework using ADS software, incorporating S-parameter analysis, stability evaluation, sensitivity studies, optimization simulations, and substrate modeling, with system stability as the primary optimization metric. Our innovative switch-based architecture separates AC and DC amplification paths, eliminating DC circuit interference in AC signal processing. We implemented the AC amplification circuit using Rogers4350B substrate and ABA-52563 RF amplifiers, optimized for high-frequency operation. Vacuum state characterization through Husimi function reconstruction validated BHD performance for quantum measurement and QRNG applications. The implemented BHD achieves: (i) 1.9 GHz bandwidth, (ii) 41.5 dB SNR at 1.75 GHz, (iii) 30 dB CMRR at 100 MHz, and (iv) $<\pm 2$ dB flatness across 1.3-1.7 GHz. This performance satisfies CV-QKD requirements for key generation rates and increases the quantum random number generator generation rate to 20.0504 Gbps.